# Heterogeneous integration of high endurance ferroelectric and piezoelectric epitaxial $BaTiO_3$ devices on Si.

Asraful Haque*, Harshal Jason D'Souza, Shubham Kumar Parate, Rama Satya Sandilya, Srinivasan Raghavan*, Pavan Nukala*.

Center for Nanoscience and Engineering, Indian Institute of Science, Bengaluru, 560012

**Abstract:** Integrating epitaxial $BaTiO_3$ (BTO) with Si is essential for leveraging its ferroelectric, piezoelectric, and nonlinear optical properties in microelectronics. Recently, heterogeneous integration approaches that involve growth of BTO on ideal substrates followed by transfer to a desired substrate show promise of achieving excellent device-quality films. However, beyond simple demonstrations of the existence of ferroelectricity, robust devices with high endurance were not yet demonstrated on Si using the latter approach. Here, using a novel two-step approach to synthesize epitaxial BTO using pulsed laser deposition (PLD) on water soluble $Sr_3Al_2O_7$ (SAO) (on $SrTiO_3$ (STO) substrates), we demonstrate successful integration of high-quality BTO capacitors on Si, with $P_r$= 7 μC/cm$^2$, $E_c$= 150 kV/cm, ferroelectric and electromechanical endurance of > $10^6$ cycles. We further address the challenge of cracking and disintegration of thicker films by first transferring a large area (5 mm x 5 mm) of the templated layer of BTO (~30 nm thick) on the desired substrate, followed by the growth of high-quality BTO on this substrate, as revealed by HRXRD and HRSTEM measurements. These templated Si substrates offer a versatile platform for integrating any epitaxial complex oxides with diverse functionalities onto any inorganic substrate.

## Introduction:

The epitaxial growth of perovskite-structure oxide thin films on silicon (Si) substrates represents a promising avenue for the integration of diverse electronic, optoelectronic, and acoustic functionalities within the established framework of low-cost Si-based integrated circuit technology [1–3]. $BaTiO_3$ (BTO) has garnered significant attention for its notable ferroelectric [4], electromechanical [5], electrooptic [6], and nonlinear optical properties [7], holding potential for application in integrated electronic and optical devices [8–10]. Nevertheless, direct epitaxial growth of BTO onto Si encounters challenges stemming from the native amorphous $SiO_x$ on Si, the chemical reactivity at the BTO/Si interface, and the substantial lattice mismatch (approximately 4.43%) between BTO and Si [11]. Consequently, various buffer layers, including

MgO, TiN, and $SrTiO_3$ (STO), and combinations of various layers, such as $LaNiO_3/CeO_2$/YSZ, were employed to facilitate the epitaxial growth of BTO onto silicon substrates [12–14].

Incorporating buffer layers on Si presents challenges related to the redox reactions and interdiffusion, as well as unintentional dead layers at the Si interface [15,16], resulting in compromised device performance. Furthermore, in Table S1, we present a comparison of crystalline qualities and ferroelectric properties obtained using various buffer layers (in the form of Full Width at Half Maximum (FWHM) of the on-axis rocking curves and PE loop characteristics), showing that epitaxial STO buffer layers obtained via molecular beam epitaxy yields the best crystalline quality.

In contrast to the monolithic growth enabled by buffer layers, heterogenous integration strategies of growing complex oxides on solvent-soluble templates [17] or using remote epitaxy on 2D materials [18], or hybrid strategies [19] are recently becoming popular. In particular, $Sr_3Al_2O_6$ (SAO) water-soluble sacrificial layer-based transfer processes are more popular than other approaches due to their ease of removal with water. Table 1 summarizes BTO membranes obtained using the SAO sacrificial layer, highlighting their applications in super-elasticity, strain-induced polarization studies, piezoelectric resonators, and photo-actuation, among others. Apart from water-soluble sacrificial layers, other solvent-soluble layers were also used to integrate epitaxial BTO layers heterogeneously. M. Sheeraz et al. [20]. demonstrated the transfer of a $SrRuO_3$ (SRO) layer using $SrCuO_2$ (SCO) sacrificial layer that dissolves in potassium iodide solution, enabling the growth of BTO on the transferred SRO template on silicon. PFM-based measurements were used to infer ferroelectricity on these samples. D. Pesquera et al. [21] demonstrated the epitaxial liftoff and transfer of BTO onto silicon and flexible substrates using a $La_{0.5}Sr_{0.5}MnO_3$ (LSMO) sacrificial layer by dissolving it in a solution of hydrochloric acid and potassium iodide. Apart from demonstrating ferroelectric switching through PE loops ($P_r < 10$ µC/cm$^2$, $E_c < 10$ kV/cm), mechanical tunability of the dielectric was also shown on these samples. In addition, L. Dai et al. [18] successfully grew on Ge substrates using remote epitaxy followed by its exfoliation and transfer onto Si to show flexoelectricity measured through conducting AFM (CAFM).

**Table 1:** BTO membranes fabricated using SAO sacrificial layer, with their properties reported in the literature.

| SAO-layered heterostructure | Measurement method employed | Property measured | Year | Ref |
|---|---|---|---|---|
| BTO/SAO//STO | PFM | The relationship between membrane thickness and crack formation are obtained. | 2020 | [22] |
| BTO/SAO//STO | PFM | Polarisation evolution in the membrane transferred onto PDMS. | 2022 | [23] |
| BTO/SRO/SAO//STO | PFM | Strain-induced polarization phenomena in flexible membranes. | 2024 | [24] |
| BTO/SAO//STO | AFM | Tunable friction in periodic wrinkles of BTO on PDMS. | 2022 | [25] |
| BTO/SAO//STO | PFM | Tip-induced in-plane ferroelectricity in wrinkled BTO. | 2022 | [26] |
| BTO/SAO//STO | PFM | BTO Ferroelectric domain wall memory on Si. | 2022 | [27] |
| BTO/SAO//$GdScO_3$ | LDV, PFM | Photoactuation in freestanding ferroelectric BTO membrane. | 2024 | [28] |
| BTO/SAO//STO | PFM, PE loop | Ferroelectric membranes are transferred onto PET. | 2020 | [17] |
| BTO/SAO//STO | PFM, in-situ SEM bending, PE loop | Demonstration of superelasticity and ultraflexibility of BTO membrane. | 2019 | [29] |
| BTO/SAO//STO | PFM, laser interferometry-based measurements. | Piezoelectric resonators with BTO membranes. | 2022 | [30] |

Water soluble epitaxial templates such as SAO are grown with good crystalline quality at very low pressures ($2x10^{-5}$ mbar) in pulsed laser deposition [31]. Device quality BTO typically requires growth at much higher pressures ($5x10^{-3}$ mbar) [13]. Combining the two processes in the same chamber would entail compromises on both the crystal quality of SAO (which in turn affects the quality of BTO) and the device performance of BTO. As a result, although these works demonstrate the existence of ferroelectricity on transferred membranes, the question of obtaining robust ferroelectric and piezoelectric device quality membranes with large endurance remains to be shown.

In this study, using pulsed laser deposition, SAO was epitaxially grown on STO at low pressure ($2x10^{-5}$ mbar), followed by the growth of epitaxial BTO in two steps: an initial 30 nm layer at low pressure ($2x10^{-5}$ mbar, LPBTO), and a subsequent layer at higher pressure ($5x10^{-3}$ mbar,

HPBTO). We show that the LPBTO (non-stoichiometric) layer is crucial to protect and preserve the crystalline quality of SAO while not compromising on the crystal and device quality of the HPBTO layer. Both the BTO layers were integrated onto silicon using a simple transfer method [32] that involved a single-step application of a polymer layer, dissolving the SAO layer, and transferring the BTO onto platinized Si. Our heterogeneously integrated BTO layer exhibits a symmetric rocking curve, with FWHM of 0.76°, similar to as-grown BTO on SAO//STO and also comparable to BTO grown on Si using MBE [33]. The high device quality of the obtained membranes transferred onto Si is demonstrated through ferroelectric and piezoelectric measurements on BTO capacitors, which show $P_r$= 7 µC/cm$^2$, $E_c$= 150 kV/cm, ferroelectric endurance of > $10^6$ cycles, and electromechanical endurance ~ $10^6$ cycles.

We further exploit our two-step growth process to obtain 5 mm x 5 mm large area membranes of BTO on Si by avoiding issues of cracking commonly observed in the transfer processes. For this, we first transfer a large area (5 mm x 5 mm) of LPBTO layer (30 nm thick) onto Si and then grow high-quality HPBTO on this templated Si substrate, as confirmed by HRXRD and HRSTEM measurements. Additionally, these templated Si substrates provide opportunities for the monolithic integration of other functional oxides, enhancing the versatility and applicability of the proposed integration technique.

**Results and discussion**

SAO film was grown by laser ablation of a polycrystalline stoichiometric SAO target (Sr:Al:O = 3:2:6) on an STO (001) substrate. The optimal conditions for achieving high-quality epitaxial SAO films were determined to be a laser fluence of 1.3 J/cm$^2$, a deposition pressure of $2\times10^{-5}$ mbar, and a substrate temperature of 850°C. Due to the hygroscopic nature of SAO, a protective BTO layer was deposited to facilitate post-growth characterization. In order to protect the crystallinity of SAO layer, BTO was grown in two pressure regimes where initially it was grown at low pressure (2x$10^{-5}$ mbar) followed by at higher pressure (5x$10^{-3}$ mbar, details in materials and methods) and subsequently compared with reference samples where BTO layer is deposited in just one step (Figure S1). The symmetric θ-2θ scan (Figure 1a) shows a (00l) set of SAO and BTO planes confirming films are oriented in the out-plane direction. The crystalline quality of the BTO and SAO films, as indicated by the rocking curve FWHM of the Bragg reflection for SAO (008) and BTO (002), is 0.51° and 0.61°, respectively, for the two-step grown BTO sample (Figure 1b, c). In contrast, for the single-step grown BTO in the reference sample, the rocking curve FWHM values are 0.96° for SAO (008) and 1.1° for BTO

(002) (Figure S1b, c). Therefore, the LPBTO is crucial to maintain the crystal quality of both SAO and BTO.

To get more insights into the structural analysis, cross-sectional FIB lamellae were prepared from BTO/SAO//STO. Notably, special care should be taken while preparing the FIB lamella with SAO at the interface, ensuring minimum exposure time between the FIB chamber and the TEM chamber. Even with proper desiccation, these lamellae containing SAO cannot be used for imaging at later stages since the SAO layer reacts with atmospheric moisture (Figure S2). The cross-sectional HAADF-STEM images of the Pt/Au/BTO/SAO//STO stack are shown in Figure 1e, and the corresponding STEM image of the BTO/SAO interface along the [001] axis of STO is shown in Figure 1f. SAO exhibits a cubic crystal structure with a space group $Pa\overline{3}$ and lattice parameter ($a_{SAO}$) of 15.263 Å, which is nearly 4 times that of STO ($a_{STO}$ of 3.905 Å). Notably, the multiplied lattice parameter of STO (4 x $a_{STO}$) equals 15.620 Å (Figure 1d), allowing for $(001)_{SAO}||(001)_{STO}$, $<100>_{SAO}||<100>_{STO}$, epitaxial relationship. The undisturbed SAO layer displays a unique rhombus-like contrast in projection along [101], a result of the ordering of Sr and Al cations along the [100] ("B-sites" in the perovskite nomenclature) [34 35]. In addition, the STEM-EDS chemical maps show homogenous layers and sharp interfaces (Figure S3). From the HAADF-STEM image of the BTO region, we extract quantitative local information about the direction of the polarization vector by measuring the displacement of the B-site cation (i.e., Ti) to the mass center of the four cations at the unit cell corners (i.e., Ba) across each unit cell in the image (Figure 1g). For BTO, the extracted polarization map reveals a strong tendency for the polarization to point along the out-of-plane direction.

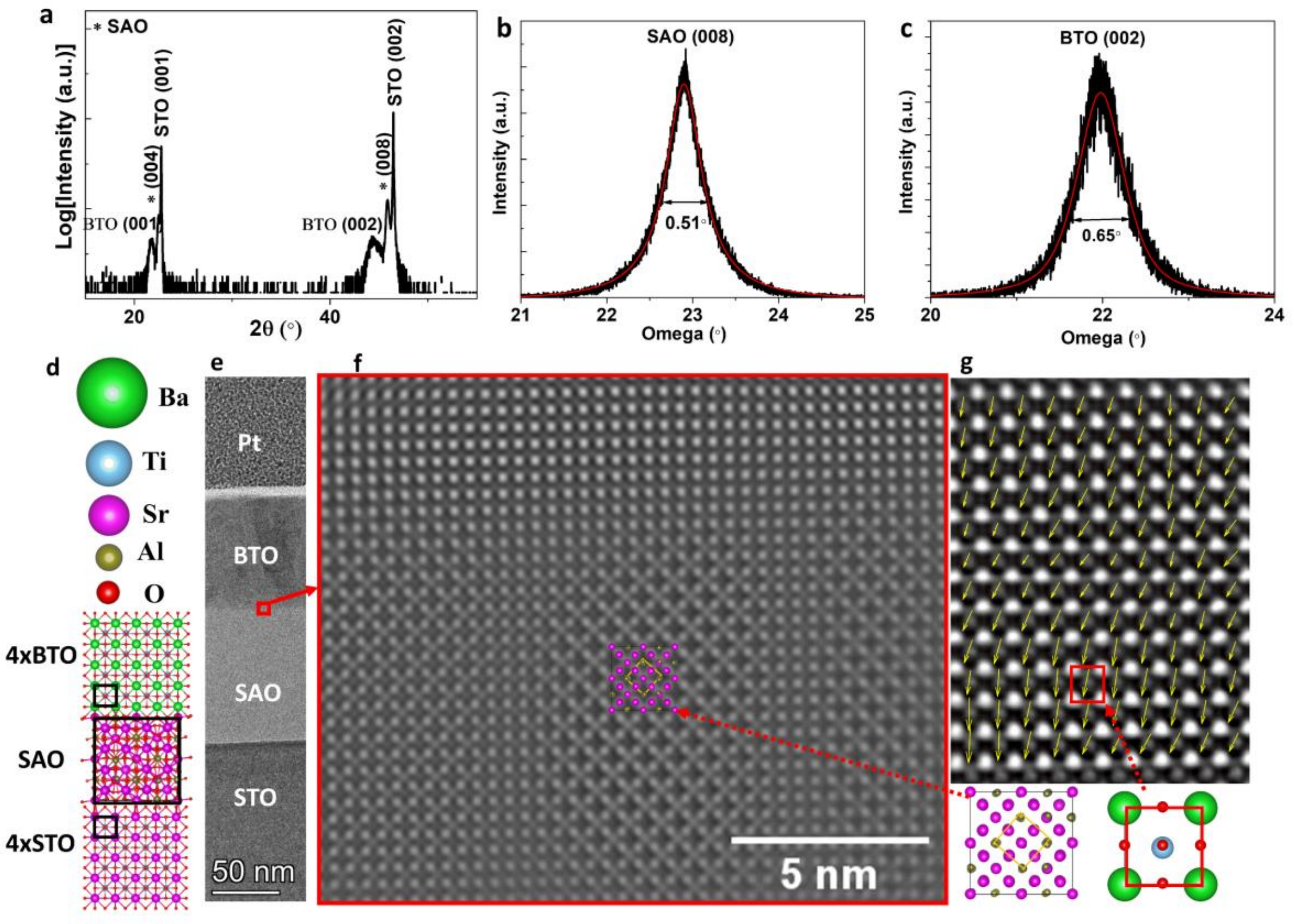

**Figure 1: Structural characterization of the BTO (grown by two steps) and SAO sacrificial layer grown on STO.** (a) X-ray diffraction θ-2θ scan of the heterostructure, (b, c) Rocking curve around BTO (002) and SAO (008) Bragg reflection. (d) A schematic of BTO/SAO thin film heterostructure on STO (001) substrate, demonstrating how four units of BTO and STO cells closely match with one SAO unit cell, enabling epitaxy. (e) The cross-sectional HAADF-STEM image of the as-grown epitaxial BTO/SAO hetero-bilayer film on the STO (001) substrate. (f) Atomic resolution HAADF-STEM images showing the crystal structure of SAO (inset, superimposed structure of SAO), (g) Ti displacement mapping in the BTO overlaid on the cross-sectional HAADF-STEM image. Ti displacement in BTO unit cells corresponds to the direction of dipole moments.

**Transfer of ferroelectric BTO membrane and its structural and electrical characterization:**

Following the successful epitaxial growth of BTO on SAO//STO heterostructure, a methodology for the exfoliation and transfer procedure was developed. Polymethyl methacrylate (PMMA) 950A2 was spin-coated at 4000 rpm onto the BTO/SAO//STO heterostructure to facilitate mechanical handling. Given PMMA's transparency, four reference dots were marked at the corners of the PMMA layer for visual aid (refer to Figure 2b). Subsequently, the PMMA/BTO/SAO//STO heterostructure was immersed in deionized (DI) water for 30-60 minutes to dissolve the sacrificial layer. The dissolution front of SAO was monitored under an optical microscope (see Figure 2c). Once SAO was completely dissolved, the PMMA/BTO layer was separated from the STO substrate by immersing the stack in water at an angle of ~45º. The floated PMMA/BTO film was then scooped with Si. To mitigate wrinkling, folds, and cracks that occur due to the potential formation of tiny channels as water drains, we dried out water from the interface between the membrane and Si substrate by heating it to 70 ºC, resulting in a clean and smooth interface while minimizing wrinkle density. Subsequently, the PMMA layer was removed by dipping it into acetone at 45 °C (Figure 2b). Through the implementation of this method, the successful transfer of millimeter-sized membranes was achieved for thin BTO layers, in this case, LPBTO (~30-100 nm in thickness). For much thicker layers, this transfer process results in membrane cracking and disintegration into pieces.

Due to the efficient dissolution of the SAO sacrificial layer by water, the integrity and alignment of the epitaxial BTO layer remain unaffected during the transfer process, as evidenced by the XRD symmetrical θ-2θ scan depicted in Figure 2e. Following the selective etching process, the (004) and (008) peaks originating from the SAO sacrificial layer were distinctly absent, while the (001) and (002) peaks from the single crystalline BTO film were evident in the transferred BTO film on Si substrate. This observation indicates that the out-of-plane crystallographic orientation of the BTO membrane remains consistent with the originally grown epitaxial BTO thin film on SAO//STO. The FWHM of the rocking curve for the (002) Bragg reflection of the transferred BTO, measured at 0.76°, closely resembles the as-grown BTO value of 0.65° (Figure 2f, Figure 1c). Furthermore, XRD φ-scan was conducted for the (103) asymmetric Bragg reflection, as illustrated in Figure 2g. The appearance of BTO (103) peaks at 90° intervals confirms the maintenance of in-plane crystallographic orientation post-transfer without any rotational domains. The mean arithmetic surface roughness of the transferred BTO membrane ($\sigma_{rms}$= 0.32 nm) closely matches that of the originally grown BTO film ($\sigma_{rms}$= 0.27 nm) on the SAO//STO, indicating a surface free of residue following the removal of PMMA (see Figure 2a, d). In a demonstration of hetero-symmetric integration, an unlikely scenario through a monolithic approach, the BTO (with cubic symmetry) membrane was transferred onto a sapphire ($Al_2O_3$ (0001)) substrate (with hexagonal symmetry). Figure 2h displays the XRD θ-2θ scan of the BTO membrane transferred onto the $Al_2O_3$ (0001) substrate. The LPBTO membrane (~30 nm) was transferred onto a TEM grid (refer to Figure 2i for the transfer process and Figure 2j for the optical image of the transferred membrane on Cu grid), and plan-view HAADF-STEM images were obtained (Figure 2k), which further confirm the high quality single crystalline nature of LPBTO layers.

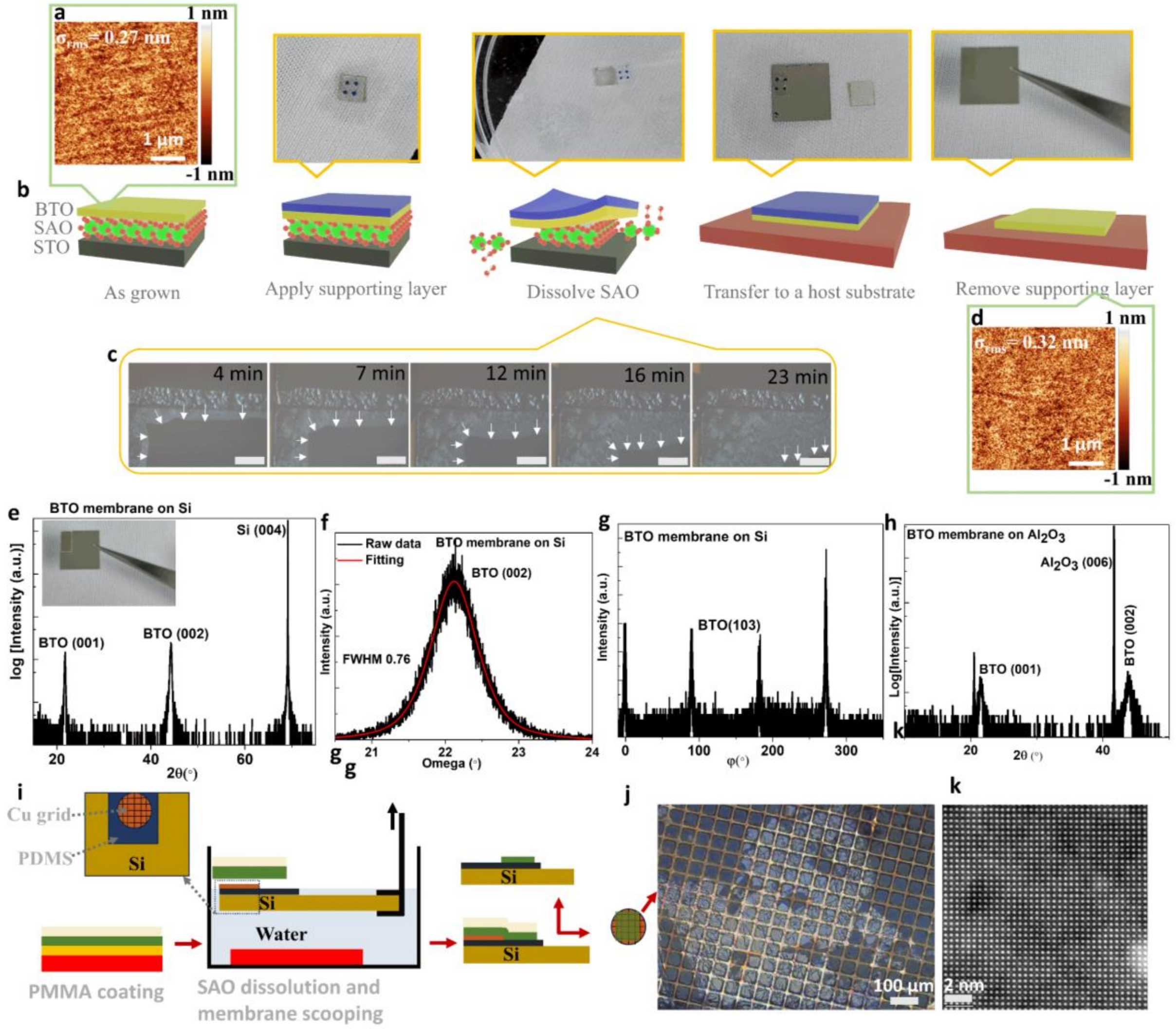

**Figure 2**: **Exfoliation, transfer, and characterization of single crystalline BTO membranes.** (a) AFM surface topography of the as-grown BTO on SAO//STO ($\sigma_{rms}$ of 0.27 nm). (b) Schematic illustration of the transfer process with corresponding optical images of the sample. The four dots in the supporting layer coated BTO/SAO//STO sample are for visual guidance during the floating and subsequent scooping process. (c) Optical image of the BTO/SAO//STO sample inside water, with white arrows indicating the dissolution front of SAO with time. (d) AFM surface topography of the transfer BTO membrane on Si after removal of the mechanical support ($\sigma_{rms}$ of 0.32 nm). (e) X-ray diffraction θ-2θ scan for the transferred

BTO membrane on Si. The inset shows an optical image of the transferred membrane on Si. (f) Rocking curve around BTO (002) Bragg reflection of transferred BTO membrane on Si. (g) φ-scan for (103) Bragg reflection of the BTO membrane on Si. (h) X-ray diffraction θ-2θ scan for the transferred BTO membrane on hexagonal sapphire substrate, illustrating epitaxial integration of cubic lattice on hexagonal lattice through layer transfer method. (i) Schematic of the BTO membrane transfer process on TEM Cu grid. A PDMS sheet was attached to the Si substrate, and a Cu grid was placed on top. After that, BTO was transferred onto the Cu grid following steps in (b), and the Cu grid was finally detached from PDMS. (j) Optical image of the 30 nm BTO membrane transferred onto Cu grid, (k) Atomic resolution plan-view HAADF-STEM image of a 30 nm thick BTO membrane.

For device measurements on the membrane, prior to transfer, Ti/Pt circular electrodes were fabricated on BTO/SAO//STO utilizing standard photolithography followed by metallization. Special care was taken to avoid any contact with water during the lithography in order to protect the SAO layer. Figure 3a illustrates an SEM image of the Pt/Ti electrode on the BTO membrane transferred onto the Pt/Si substrate (a schematic of the device cross-section is in the inset). Instantaneous current vs. voltage and polarization versus voltage curves for the BTO membrane, performed at 25°C and 1 KHz, clearly show the polarization switching peaks (Figure 3b) and ensuing ferroelectric hysteresis (Figure 3c). The remanent polarization ($P_r$) and coercive field ($E_c$) values were determined to be ~7 μC/cm$^2$ and 150 kV/cm, respectively. $P_r$ is smaller than the ideal bulk single crystalline values of 25 μC/cm$^2$ [17,36], but are comparable to other works on BTO membranes [37]. Furthermore, Typical $E_c$ values for BTO are ~100 kV/cm [13] in thin films, and our values are larger than these. These results point towards the unoptimized nature of the BTO and transferred substrate interfaces and motivate further studies to improve these interfaces [38]. Polarization pinning effects at the interfaces may lead to a reduction in $P_r$ as well as an increase in $E_c$.

Next, we tested the endurance of these devices by applying cyclic rectangular voltage waveforms at 1 MHz and reading the polarization through PUND measurements every few cycles. Our devices did not show any indication of wake-up or fatigue, exhibiting stability exceeding $10^6$ cycles, as shown in Figure 3d. Polarization and current loops after endurance testing (>$10^6$ cycles) do not show any significant differences from the uncycled state (Figure S4).

Next, we measured the electrostrain response of our devices by subjecting the device to a large signal AC voltage and measuring the out-of-plane displacement through a laser-doppler vibrometer (double laser). Figure 3e presents butterfly loops of voltage vs. time-averaged displacement response over 70000 sinusoidal voltage cycles ($V_{max}$=20 V, Frequency=6 kHz). A phase lag is observed between the applied voltage and the resultant displacement response, leading to a non-zero displacement at zero voltage, as depicted in Figure 3e. The upturns of the butterfly loops coincide with the coercive voltage, measured from peak positions of the instantaneous current vs voltage curves, typical of the electrostrain response of a ferroelectric material. An effective piezoelectric coefficient ($d_{33}^{effective}$) ~11 pm/V was extracted from these measurements for devices with both 50 µm and 100 µm electrode diameters (Figure 3e, Figure S5). In general, it may be noted that the exact value of $d_{33}^{effective}$ will depend on the electrode size [39] and film thickness [40]. However, these values are in the same order as $d_{33}^{effective}$ reported on BTO films monolithically grown on various single crystal oxide substrates [40].

Next, we cycled these devices with $V_{max}$= 20 V (E= 600 kV/cm) through the polarization switching and measured the corresponding electromechanical strain response. This accounts for strain cycling with a max amplitude of electrostrain ~0.1%. Our devices start fatiguing only after $10^6$ cycles, which shows that these are very robust for micro and nanoelectromechanical applications [5](Figure 3f).

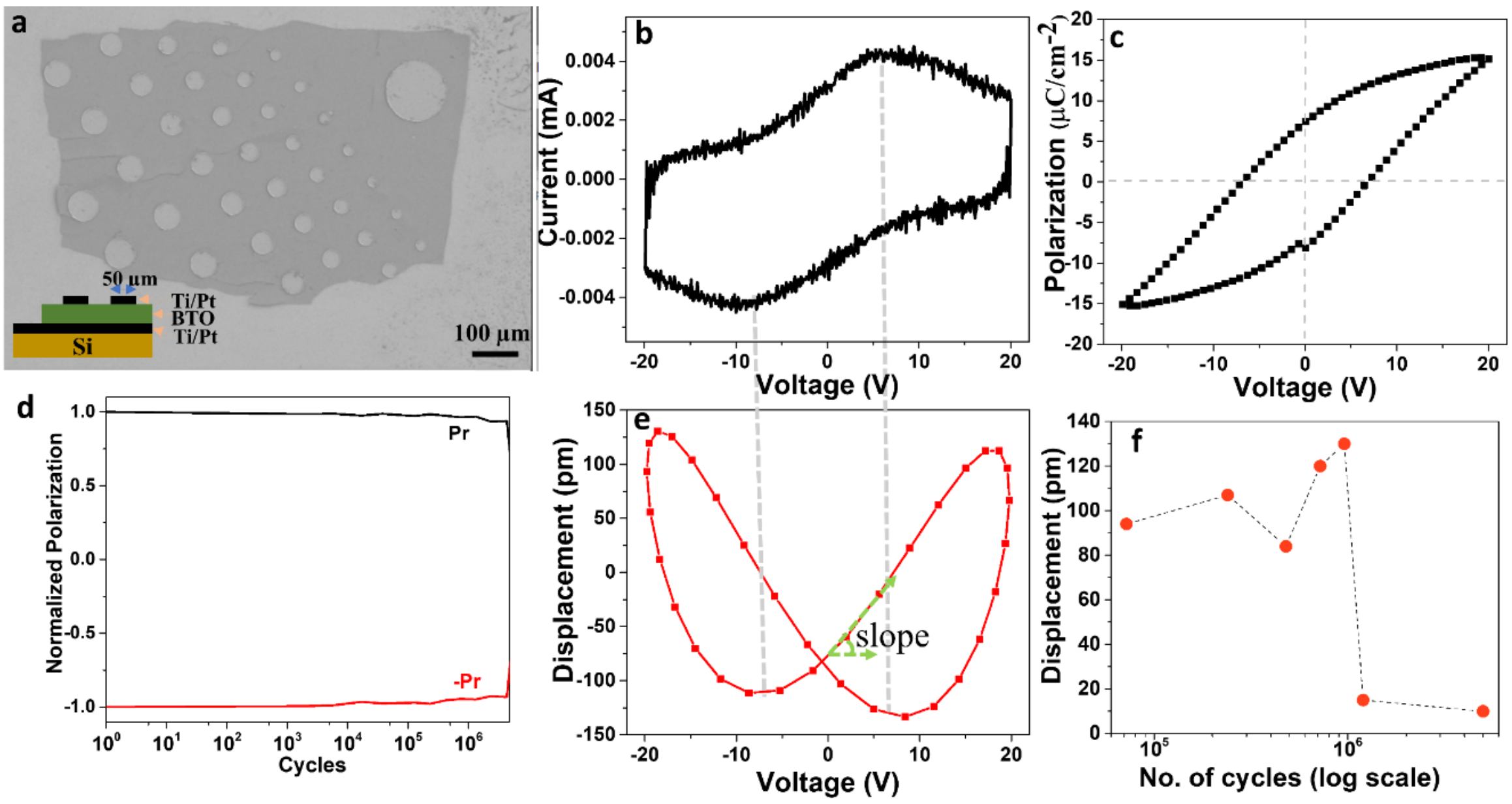

**Figure 3: Ferroelectric and Piezoelectric properties of BTO membrane-based capacitors heterogeneously integrated on Si**. (a) SEM image of BTO membrane with top Pt electrode

transferred onto Pt/Si substrate. The inset shows a schematic of the cross-section with devices. (inset) schematic of device cross-section. (b) Instantaneous current (I) vs Voltage (V) plot of the BTO membrane on Pt/Si substrate, measured at 1000 Hz. (c) Current (I) vs. voltage loop for the BTO membrane on Pt/Si substrate. (d) Normalized remanent polarization as a function of the number of switching cycles at 1 MHz (endurance test). (e) Average mechanical displacement vs voltage response. (g) Mechanical displacement vs no. of cycles response of a representative device, tested for electromechanical endurance at at 20V peak voltage and 6 kHz.

**Templated Si substrate and epitaxial growth on it:**

It is crucial to note that electromechanical devices employ the usage of thicker films (t >200 nm). Thicker films disintegrate during transfer, and thinner films are better for large-area transfers. We circumvent the issue of cracking in the thicker membranes by first transferring a 5 mm x 5 mm (large scale) LPBTO membrane (~30 nm) onto Si. Subsequently, we grow HPBTO epitaxially on this template, resembling conditions of homoepitaxy.

Reciprocal space mapping (RSM) of the as-grown stack (STO//SAO/LPBTO, schematic in Figure 4a) taken near STO (002) Bragg reflection shows fully strained LPBTO and SAO layers on STO, indicating coherent interfaces (Figure 4d). Figure 4e shows the SEM image of a large area of LPBTO transferred onto Si (schematic cross-section in Fig 4b), which will be referred to from now on as templated Si. The surface roughness of the templated Si is measured from AFM to be $\sigma_{rms}$ = 0.38 nm, illustrating smooth surfaces (Figure 4f). RSM of LPBTO (002) Bragg reflection, as shown in Figure 4g, shows a slight shift in 2θ towards a higher angle due to film relaxation post-releasing.

For the HPBTO grown on templated Si (schematic in Figure 4c), the RSM around (002) Bragg reflection is shown in Figure 4h. The FWHM of the on-axis rocking curve, extracted from RSM, is 0.7°, showing a slight degradation in crystalline quality as compared to that of the LPBTO on the template (FWHM=0.5°). The surface of the deposited large area (5 mm x 5mm, crack-free) HPBTO is also very smooth ($\sigma_{rms}$ =0.41 nm), as shown in Figure 4i.

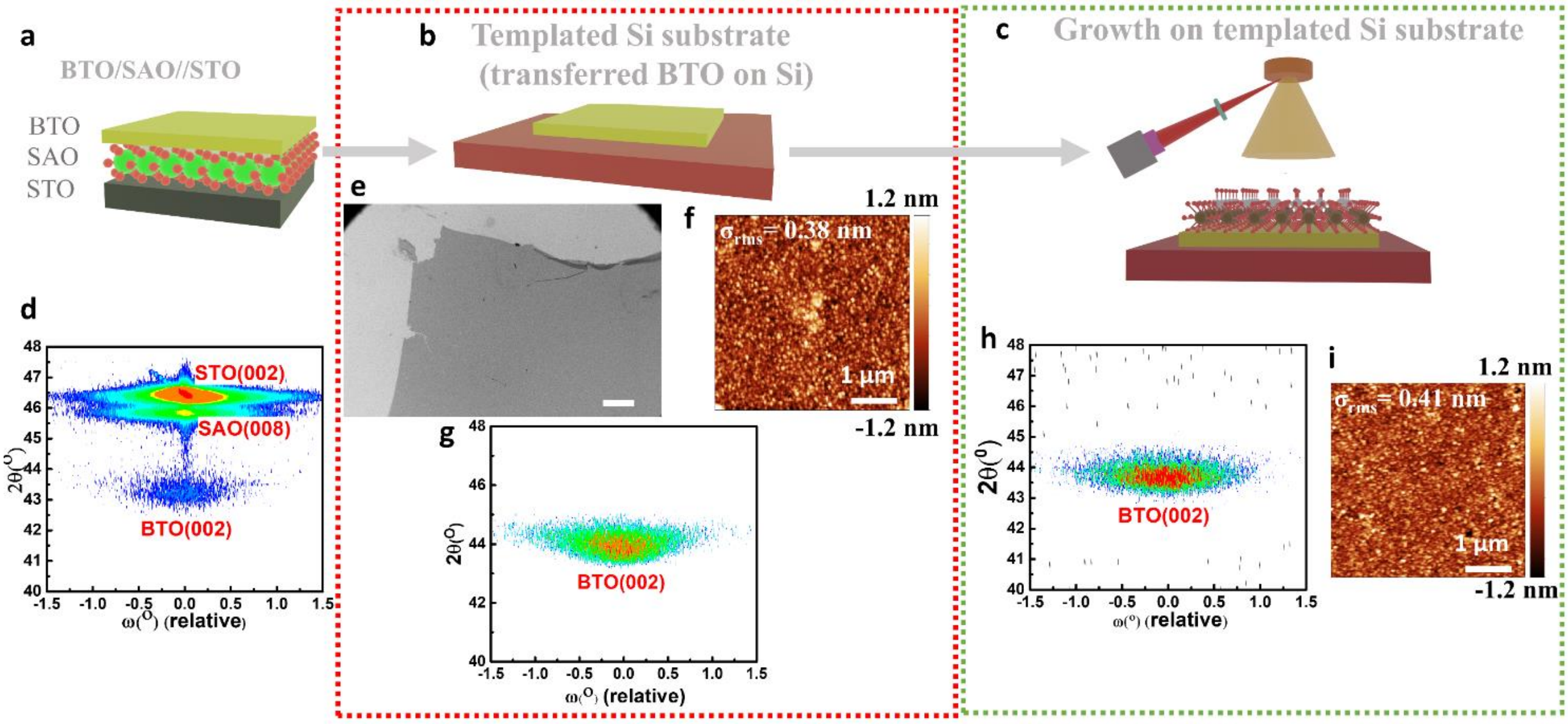

**Figure 4: Templating Si with LPBTO and subsequent epitaxial growth of HPBTO.** (a, b) Shows the schematic of the transfer process of LPBTO onto Si, creating a templated Si. (c) Schematic of PLD-assisted growth of HPBTO on templated Si. (d) Reciprocal space mapping (RSM) around (002) Bragg diffraction of STO. (e) SEM micrograph showing large area transfer of LPBTO membrane on Si (scale bar 500 μm), (f) AFM surface topography ($\sigma_{rms}$ of 0.38 nm) of the templated Si from (e). (g) RSM of the BTO (002) Bragg diffraction on the templated Si (shown in b, e), (h) RSM of BTO (002) Bragg diffraction on HPBTO/Si template, (f) AFM surface topography of HPBTO ($\sigma_{rms}$ of 0.41 nm).

To study the chemical composition and microstructure of the interfaces, cross-sections of these samples were analyzed using high-angle annular dark-field scanning transmission electron microscopy (HAADF-STEM) imaging (see Figure 5a-d), correlated through energy-dispersive X-ray spectroscopy (STEM-EDX) (see Figure S6). Figure S6 shows that there is no interdiffusion of elements at the BTO/Si interface, and Figures 5a and S6 show well-defined LPBTO/Si and LPBTO/HPBTO interfaces over large scale length (> 2μm), a testimony of good transfer of the LPBTO film onto Si resulting in a robust template.

The average out-of-plane (c) and in-plane lattice parameter (a) of LPBTO estimated from real space HAADF-STEM images are 4.12±0.01 A° and 3.94±0.002 A°, respectively. The larger c and smaller a lattice parameters with respect to the bulk (c=4.04 A°, a=3.99 A°) clearly indicate the defective nature of the strain-relieved LPBTO film. On this template, the deposition of stoichiometric $BaTiO_3$ will not be homoepitaxial. The template offers in-plane compressive stress to the HPBTO, which, beyond a certain thickness will relax, forming various defects. On the HPBTO, we measured the c-parameter to be 4.02±0.01 A° (Figure 5e) and the a-parameter to be 3.98±0.02 to 4.07±0.04A. We identify several defects, including antiphase boundaries, low-angle grain boundaries (and dislocation arrays, Figure 5f-h), and domain boundaries.

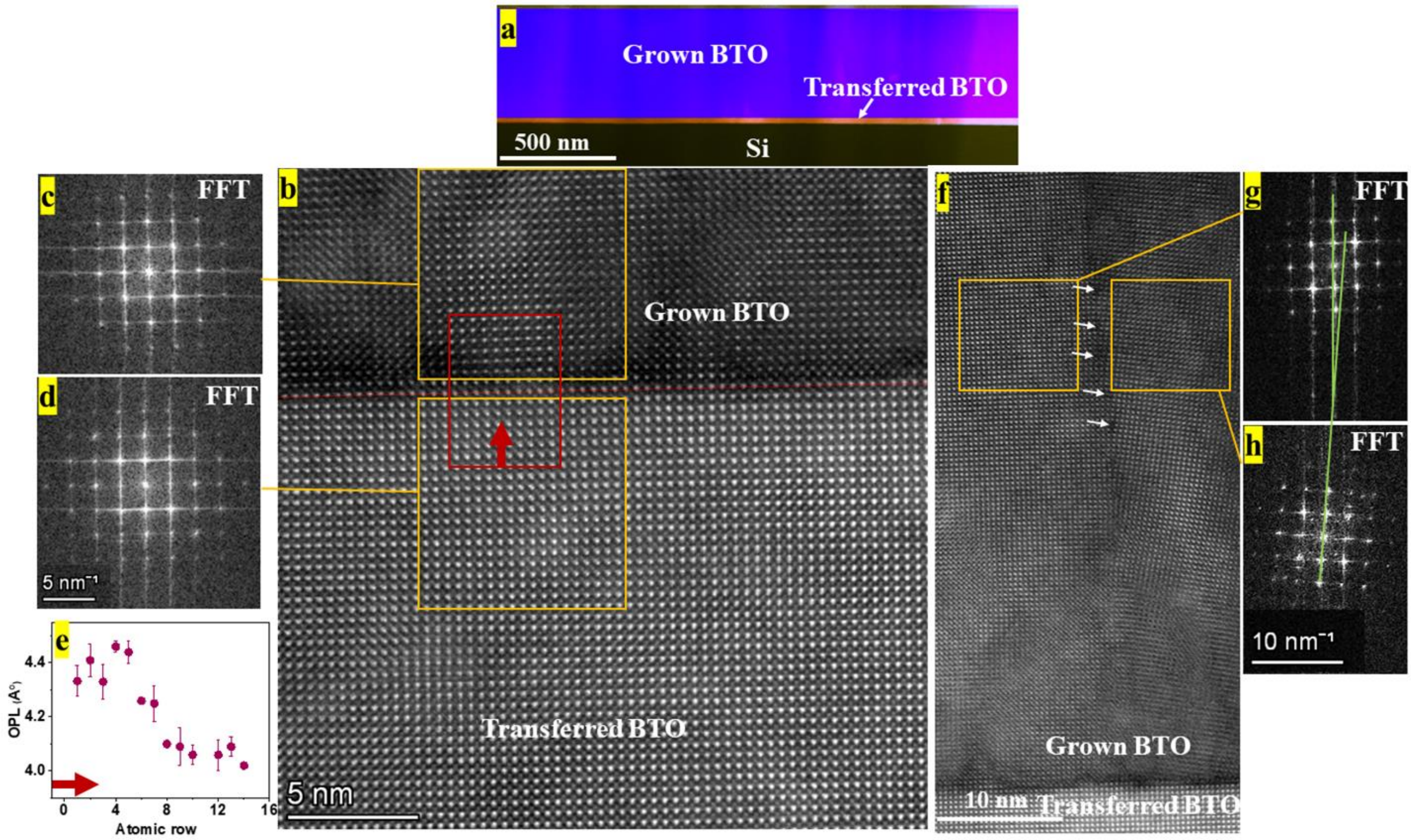

**Figure 5: Atomic and microstructural features of the HPBTO films integrated on Si-LPBTO templates.** (a) Large area HAADF-STEM image (shown in false colour, field of view=2.2 µm x 0.6 µm) of the cross-section of HPBTO grown on transferred LPBTO on Si (templated Si). (b) Atomic resolution HAADF-STEM image showing the sharp interface between grown BTO (HPBTO) on transferred BTO on Si (LPBTO). (c, d) corresponding FFTs of regions marked in (b). (e)The average out-of-plane lattice parameter as a function of atomic rows is denoted by the arrow (corresponding to the arrow shown in b). (f) HAADF-STEM image of HPBTO layer showing microstructural defects such as low angle grain boundaries (and arrays of dislocation that form them indicated by white arrows) and (g, h) FFT of the regions (across two grains), shown in (f).

To understand the direction of polarization and the nature of domain boundaries, the relative displacement of Ti in relation to Ba was mapped across various regions in the sample (Figure 5). The polarization map from the LPBTO template shows that it is mostly in the out-of-plane direction (Figure 5b), as evident from the polar plot of polarization vectors (Figure 5c) collected across various images. In HPBTO, the interfacial region close to LPBTO retains the out-of-plane polarization. However, away from this interface, we identify regions that have out-of-plane and in-plane polarizations. The domain boundaries can also be identified in low magnification HAADF-STEM images as regions separating brighter from darker contrast arising due to channeling effects. Thus, our sample is replete with ferroelastic 90° domain boundaries, suggesting strain relaxation effects in thicker HPBTO layers.

.

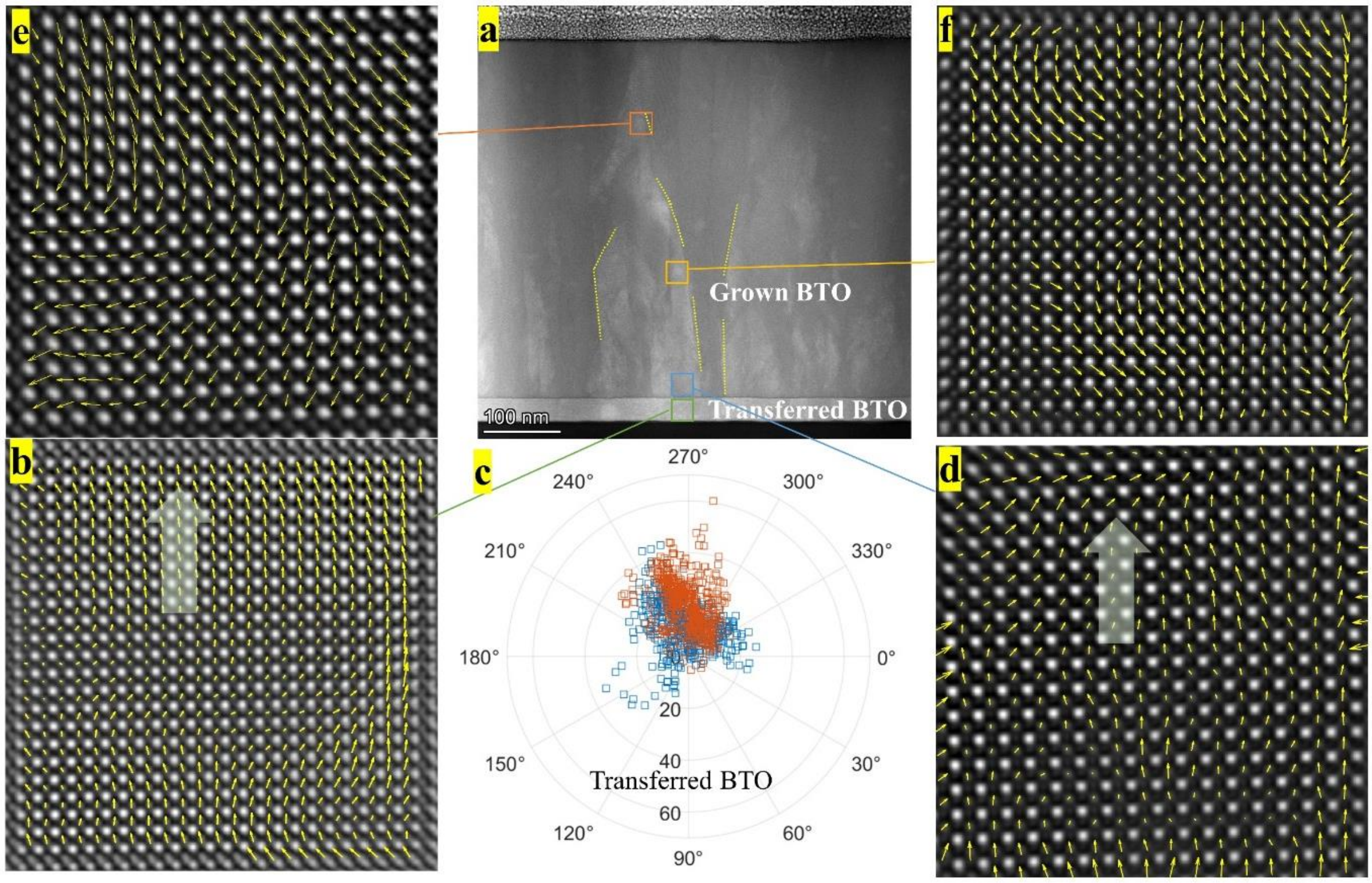

**Figure 6: Polarization mapping of LPBTO and various regions in HPBTO.** (a) Cross-sectional HAADF STEM image of HPBTO grown on templated Si. Ferroelastic domain boundaries are marked by yellow lines. (b) Polarisation map from the LPBTO region on Si shows that it is mostly in the out-of-plane direction, (c) Polar plots of polarisation vector collected across two different regions (labeled in red and blue colour) in LPBTO confirms the out-of-plane polarization. (d) The polarisation mapping in the region close to LPBTO retains the out-of-plane polarisation. (e, f) Polarisation mapping in the region away from the interface is marked in (a). In (e), the 90° domain boundary is identified as the region that separates down and diagonal polarizations with regions with horizontal polarizations.

**Conclusions:**

In conclusion, we present a novel buffer layer-free method for the heterogeneous integration of robust ferroelectric BTO on the Si CMOS platform. We employed a two-step growth of BTO on SAO//STO in order to leverage its ferroelectric property without degrading the crystallinity of the SAO layer during PLD. Subsequently, the BTO layer is transferred onto Si using simple steps. Comprehensive measurements confirm the presence of robust ferroelectric with $P_r$= 7 μC/cm$^2$, $E_c$= 150 kV/cm, and ferroelectric and electromechanical endurance of > $10^6$ cycles in the transferred BTO on platinized Si. This heterogeneous integration method also facilitates the growth of cubic BTO onto even difficult substrates using monolithic integration, such as hexagonal sapphire.

This process also allows to propose and demonstrate large area growth (5 mm x 5 mm) of stoichiometric BTO (generalizable to other oxides) on Si templated with epitaxial and defective BTO. This approach addresses the challenge of large-area transfer of thicker BTO (and other complex oxide) layers, which have a tendency to crack and disintegrate into pieces.

Our strategy to release and transfer freestanding epitaxial ferroelectric BTO membranes offers an unprecedented platform to integrate emergent phenomena in epitaxial oxide heterostructures with current state-of-the-art Si-based technology, such as integrated electronics [2] and photonics [6]. For instance, our epitaxial BTO with clean interfaces on silicon could enable the development of optical modulators with low losses directly integrated into silicon, circumventing the need for a lossy buffer layer like TiN at the interface.

**Materials and Methods:**

**Preparation of $Sr_3Al_2O_6$ (SAO) and $BaTiO_3$ (BTO) targets:**

Polycrystalline $Sr_3Al_2O_6$ target was prepared through hand grinding stoichiometric amounts of $SrCO_3$ and $Al_2O_3$ powder for several hours, followed by calcination at 1200 °C for 10 hours (5 °C/min ramp rate) under ambient conditions. This grinding and calcination process was repeated multiple times to facilitate the chemical reaction:

$3SrCO_3 + Al_2O_3 \rightarrow Sr_3Al_2O_6 + CO_2\ (\uparrow)$

Calcined powders were then pelletized and sintered at 1350 °C for 24 hours. Due to its hygroscopic nature, the SAO target was always stored in a vacuum of $10^{-7}$ mbar inside the PLD chamber. Similarly, the BTO target was fabricated using BTO nano-powders, pelletized, and sintered at 1350 °C for 12 hours.

**Epitaxial thin film growth:**

Thin films were grown using pulsed laser deposition (PLD) utilizing an excimer laser with 248 nm wavelength. Sacrificial layers of SAO was grown on STO substrates at a temperature of 850 °C, $P_{O2}$ of $2x10^{-5}$ mbar and with a laser fluence of 1.5 J/cm$^2$. LPBTO films were grown at 800 °C with a $P_{O2}$ of $2x10^{-5}$ mbar, whereas HPBTO at $5x10^{-3}$ mbar with a laser fluence of 0.9 J/cm$^2$. Reflection high energy electron diffraction (RHEED) was employed to monitor BTO growth on the templated substrate.

**Transfer and characterization of freestanding membranes:**

PMMA was spin-coated on BTO/SAO layers deposited on STO (001) substrates. These structures were then immersed in room-temperature deionized water to dissolve the SAO buffer layer. Subsequently, the PMMA-supported membranes were detached and transferred onto Si substrates. Notably, the membranes remained attached to the substrate even after the PMMA layers were dissolved with acetone. Furthermore, AFM images were taken using an Asylum AFM (MFP-3D Origin) operating in tapping mode, and X-ray diffraction data were obtained using a monochromate Cu-Kα1 source on a Rigaku four-circle machine.

**STEM imaging and analysis:**

The cross-sectional FIB lamella for TEM analysis was prepared using a focused ion beam (Model: FEI, Scios2). It was then examined using a double aberration-corrected Themis microscope operated at 300 kV, as well as a non-aberration-corrected Themis microscope also operated at 300 kV, which was equipped with a chemi-STEM EDS system. STEM-EDS analysis was conducted on the non-aberration-corrected Themis microscope, and data collection continued until enough counts (SNR>5) were obtained from binned pixels.

**Polarization mapping analysis:**

Atomic resolution HAADF-STEM imaging was employed to analyze Ti displacement within each unit cell. Initially, a meticulously chosen distortion-free lattice image was utilized to estimate and quantify the displacement. Following this selection, the image underwent Bragg filtering using a Digital Micrograph to generate the final displacement map. Subsequently, an analysis of Ti displacement away from the center of mass of the Ba (A-site) unit cell was conducted using AutoMap and Temul-toolkit. The resulting displacements were visually depicted on the image using arrows, which indicated both the magnitude and direction of displacement. This method of Ti displacement analysis serves as an effective means for estimating the unit cell dipole moment of BTO.

**Device Fabrication:**

Optical lithography (utilizing Heidelberg uPG 501 direct writer) was employed to define the circular electrical contact patterns of 50 μm diameter. Subsequently, Ti/Pt (10/100 nm) contacts were deposited through D.C. sputtering.

**Ferroelectric and electromechanical response:**

Using the top-bottom electrode, ferroelectric polarization loops, current loops and ferroelectric fatigue measurement at room temperature were carried out using a Radiant technology precision multiferroic ferroelectric tester. Laser Doppler vibrometer (LDV) (model: MSA 500) equipped with 532 nm reference and probe lasers was used to study the out-of-plane electromechanical response of the BTO film in a double beam interferometry mode. Displacement responses were obtained over 30-40 ms as a function of time and subsequently averaged.

## ASSOCIATED CONTENT

**Supporting Information**.

## AUTHOR INFORMATION

**Corresponding Author**

*Srinivasan Raghavan, Email: sraghavan@iisc.ac.in

*Pavan Nukala, Email: pnukala@iisc.ac.in

*Asraful Haque, Email: asrafulhaque@iisc.ac.in

**Present Addresses**

***Center for Nanoscience and Engineering, Indian Institute of Science, Bengaluru, India, 560012**

**Notes**

The authors declare no competing financial interest.

**ACKNOWLEDGMENT**

This work was partly carried out at the Micro and Nano Characterization Facility (MNCF) and National Nanofabrication Center (NNfC) located at CeNSE, IISc Bengaluru, funded under Grant DST/NM/NNetRA/2018(G)-IISc, NIEIN, and Ministry of Human Resource and Development, Government of India and benefitted from all the help and support from the staff. We further acknowledge the advanced facility for microscopy and microanalysis (AFMM) facilities, IISc for all the high-resolution microscopy presented in this work. P.N. acknowledges Start-up grant from IISc, Infosys Young Researcher award, and DST-core research grant CRG/2022/0003506. A.H. would like to thank Dr. Mehak Mehta and Prof Susobhan Avasthi for access to high-temperature PLD. AH also thanks Dr Varun Harbola for the insightful discussions.

# Supporting Information.

## Heterogeneous integration of high endurance ferroelectric and piezoelectric epitaxial $BaTiO_3$ devices on Si.

Asraful Haque*, Harshal Jason D'Souza, Shubham Kumar Parate, Rama Satya Sandilya, Srinivasan Raghavan*, Pavan Nukala*

Center for Nanoscience and Engineering, Indian Institute of Science, Bengaluru, 560012

**Table 1:** From the literature, ferroelectricity in BTO was grown on Si and a single crystalline substrate.

| BTO heterostructure | Growth method | Epi or poly-crystalline | FWHM BTO (°) (002) | Remnant polarization ($\mu C/cm^2$) | Coercive field (kV/cm) | Ref |
|---|---|---|---|---|---|---|
| BTO/LNO/CeO2/YSZ/Si | PLD | | | 3-11 | 70-160 | [3] |
| BTO/LNO/STO//Si | STO by MBE and BTO, LNO by PLD | yes | 0.84 | 6 | 60 | [4] |
| Pt/BTO/SRO/MgO/TiN//Si | PLD | yes | 1.4 | 4-5 | 8.5 | [5] |
| TiN/BTO/TiN//Si | PLD | yes | 1.3 | No PE loop | | [6] |
| BTO/LNO//Si | Sputter | poly | | 2 | 80 | [7] |
| BTO/Pt/Ti/TiOx//Si | Sol-gel | poly | | 2 | 27 | [8] |
| BTO/Pt/SiO2/Si | Sol-gel | poly | | 1 | 90 | [9] |
| Single crystalline BTO | | | | 24 | 1.5 | [9] |
| Ceramic BTO | | | | 8 | 3 | [9] |
| BTO/STO/Si | BTO: PLD, STO: MBE | | | CV shows a ferroelectric signature. | | [10] |
| BTO/STO/Si | All MBE | yes | 0.7 | No PE loop is shown | | [10] |
| BTO/LSCO//MgO | PLD | yes | 0.7 | 14 | 65 | [11] |
| BTO/SRO//STO | PLD | yes | | 7.3 | 29.5 | [12] |
| BTO/SRO/STO | PLD | Yes | | 15.43 | 350 | [13] |
| BTO (300nm)/SRO/STO | PLD | Yes | 0.4 | 10 | 100 | [14] |
| SRO/BTO(5-30nm)/SRO/STO | PLD | yes | | 36 (30nm film) | | [15] |
| SRO/BTO/SRO//STO | Sputtering | Yes | 0.2 (20nm)-1(>50 nm film) | 33.5-38.5 (20-80nm) | 583.3 (for 12nm film) | [16] |
| BTO/SRO/DyScO3(110), GdScO3(110), NdScO3(110) | PLD | yes | NM | NM | NM | [17] |
| BTO/SRO/GdScO3(110) | PLD | yes | | 30 | 100 | [18] |
| BTO/SRO/GdScO3 or BTO/SRO/STO(MBE)//Si | PLD | yes | 0.05-0.07 (on Si not given) | 20-30 | < 10 | [19] |

LSCO: $LaSr_{0.5}Co_{0.5}O_3$, data not mentioned denoted as NM

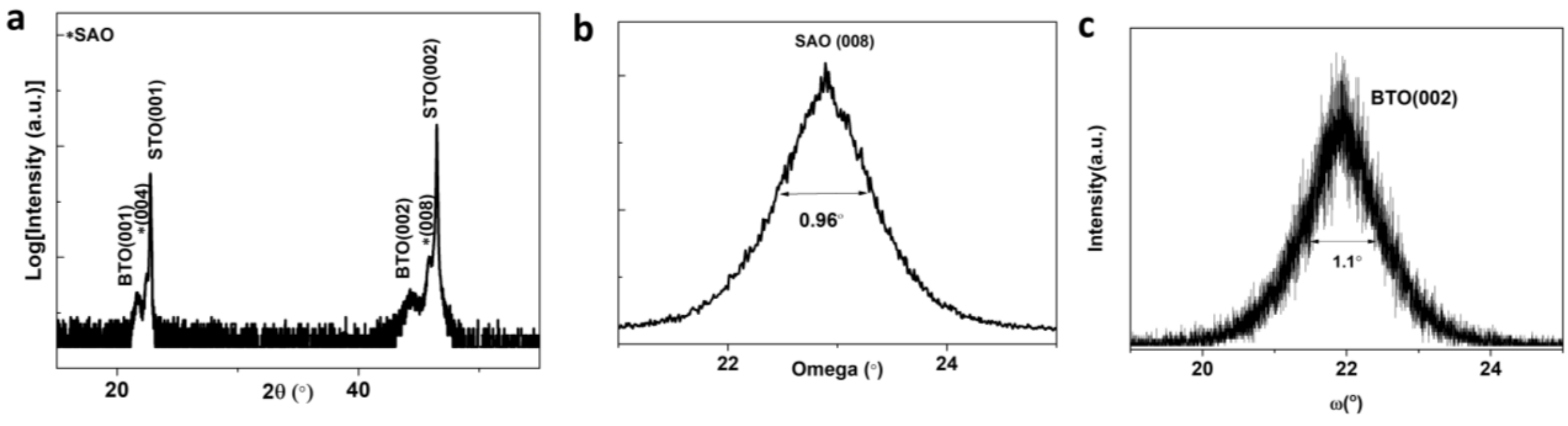

**Figure S1:** Structural characterization of the BTO (grown using a single step, reference sample) and SAO sacrificial layer grown on STO. For BTO/SAO//STO (a) X-Ray diffraction θ-2θ scan of the heterostructure, (b, c) Rocking curve around BTO (002) and SAO (008) Bragg reflection.

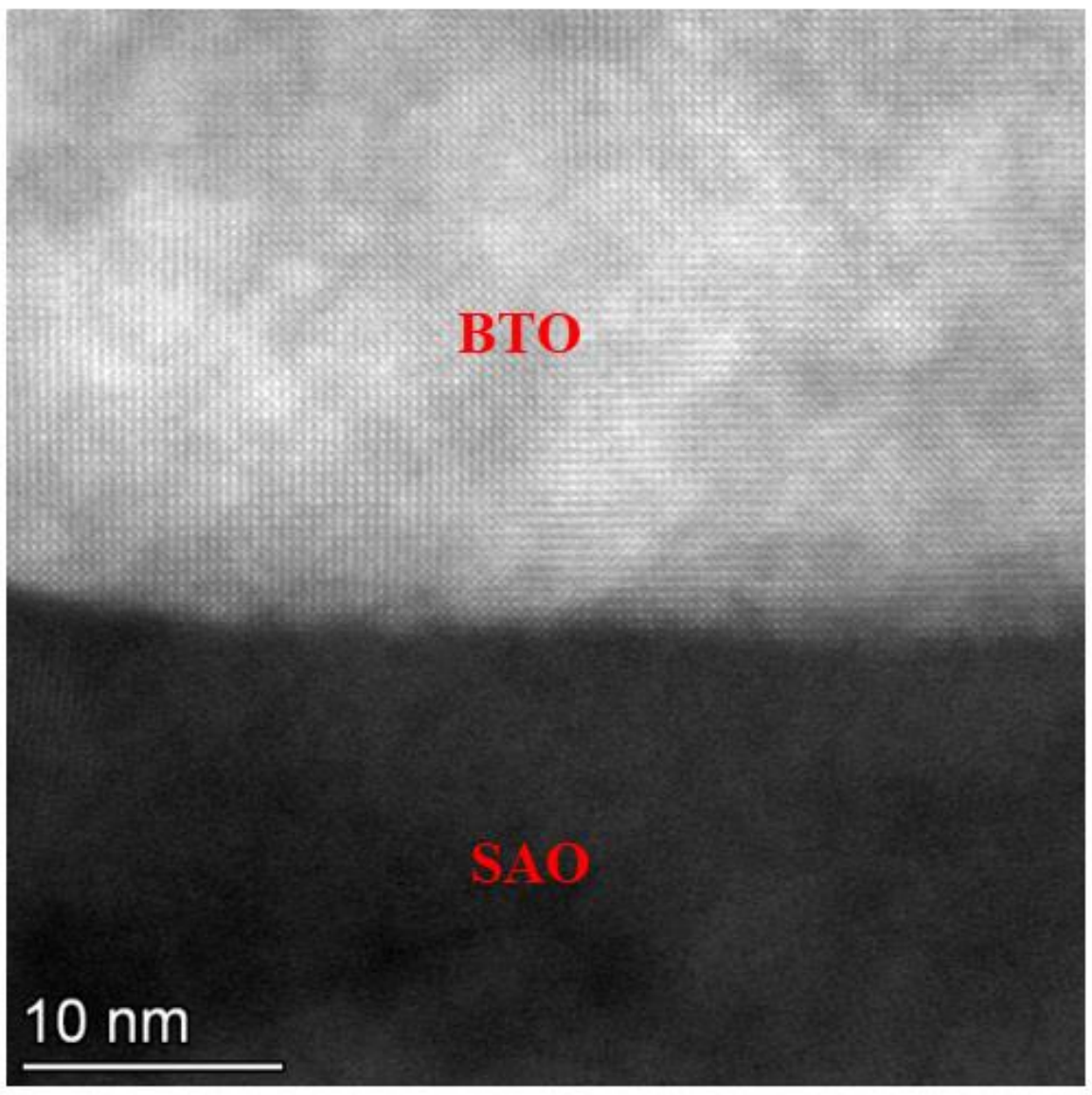

**Figure S2:** Atomic resolution HAADF STEM image of the BTO/SAO interface after the lamellae are desiccated for a day. SAO layer reacted with the atmosphere; however, an ordered lattice structure can still be seen in the BTO layer.

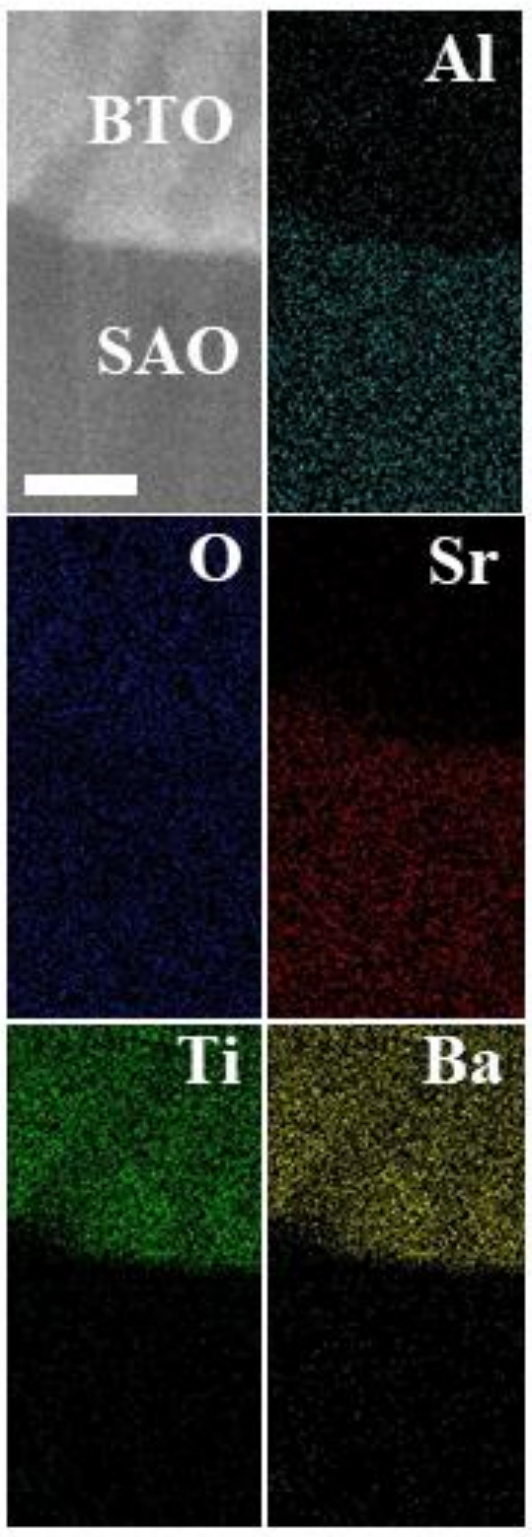

**Figure S3:** STEM-EDS shows the elemental mapping of Sr, Ti, O, Al, and Ba in the SAO and BTO regions in Pt/Au/BTO/SAO//STO.

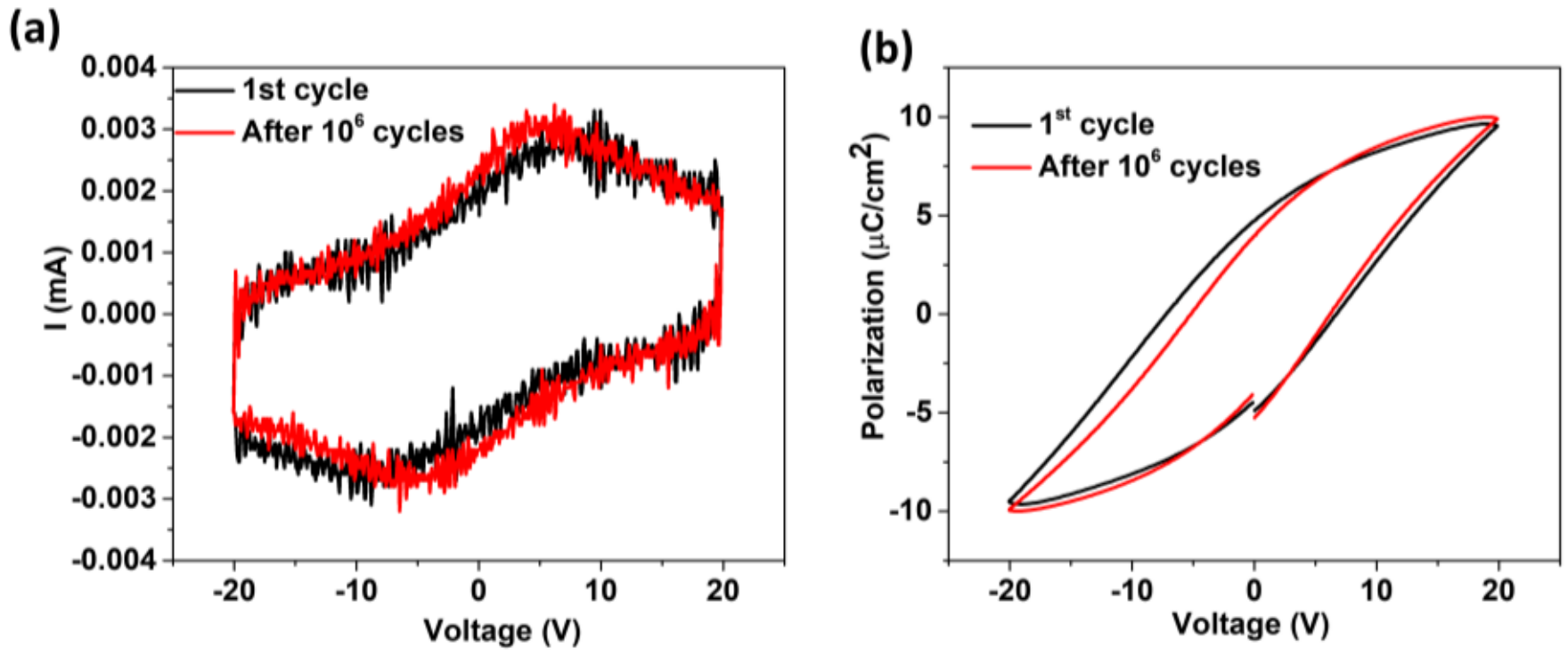

**Figure S4:** (a) Polarization vs voltage plot and (b) Instantaneous current vs voltage loop of BTO membrane (measured at 1 KHz) on Pt/Si before and after $10^6$ fatigue cycles testing at 1MHz.

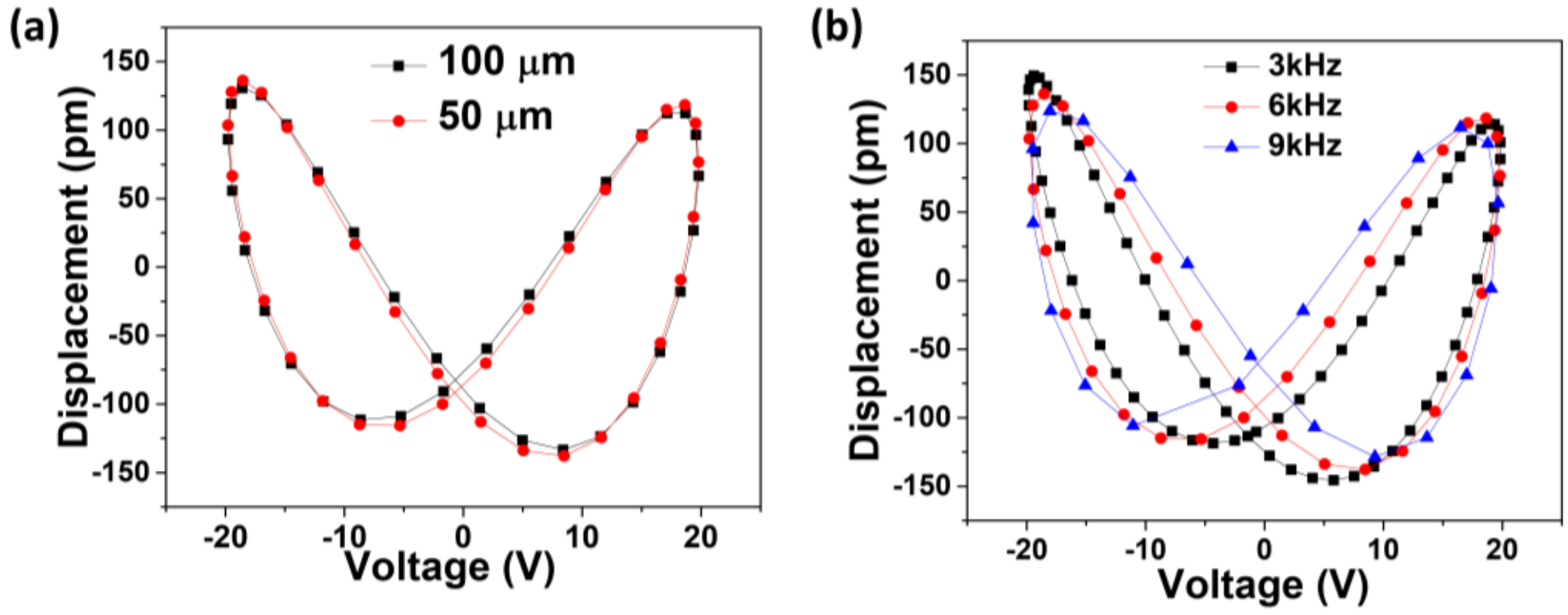

**Figure S5:** Average displacement vs voltage response of BTO membrane on Pt/Si using (a) 50 μm and 100 μm top electrodes and (b) at different frequencies using 50 μm top electrode.

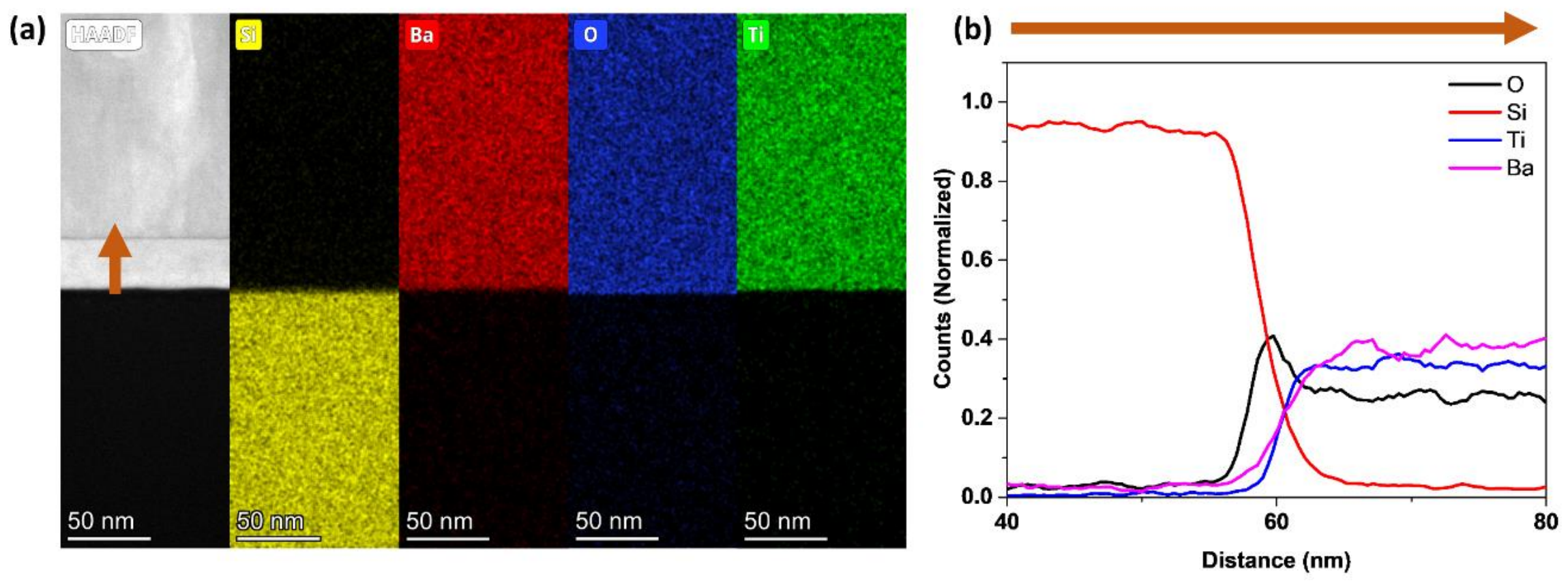

**Figure S6:** (a) STEM EDX from a cross-section of the grown BTO on the templated Si substrate. (b) Ba, Ti, Si, and O concentration profiles varied across the interfaces.